\documentclass{iau_FM}
\usepackage{graphicx}
\usepackage{natbib}
\usepackage{amssymb}
\usepackage{color}
\usepackage[colorlinks=true, linkcolor=blue, citecolor=blue, urlcolor=blue, breaklinks=true]{hyperref}
\usepackage[french, british]{babel}

\newcommand{\dd}{\mathrm{d}}

\title[Magnetising the Cosmic Web during Reionisation] 
{Magnetising the Cosmic Web\\ during Reionisation}

\author[Mathieu Langer \& Jean-Baptiste Durrive]   
{Mathieu Langer$^1$
 \and Jean-Baptiste Durrive$^2$}

\affiliation{$^1$Institut d'Astrophysique Spatiale, CNRS, UMR 8617, Univ. Paris-Sud, Universit\'e Paris-Saclay, b\^at. 121, 91405 Orsay Cedex, France\\ email: {\tt mathieu.langer@ias.u-psud.fr} \\[\affilskip]
$^2$Department of Physics and Astrophysics, Nagoya University, Nagoya 464-8602, Japan \\present email: {\tt jdurrive@irap.omp.eu}}

\pubyear{2018}
\jname{Astronomy in Focus}
\editors{Teresa Lago, ed.}
\begin{document}

\maketitle

\begin{abstract}
 Evidence repeatedly suggests that cosmological sheets, filaments and voids may be substantially magnetised today. The origin of magnetic fields in the intergalactic medium is however currently uncertain. We discuss a magnetogenesis mechanism based on the exchange of momentum between hard photons and electrons in an inhomogeneous intergalactic medium. Operating near ionising sources during the epoch of reionisation, it is capable of generating magnetic seeds of relevant strengths over scales comparable to the distance between ionising sources. Furthermore, when the contributions of all ionising sources and the distribution of gas inhomogeneities are taken into account, it leads,  by the end of reionisation, to a level of magnetisation that may account for the current magnetic fields strengths in the cosmic web.
\end{abstract}

\firstsection 
\section{Introduction}

The Universe seems to be magnetised virtually on all scales.
The origin of the cosmological magnetic fields in particular remains unsettled, despite the many models that have
been proposed  \citep[see][]{Kulsrud08, Ryu2012, Widrow2012, Durrer2013, Subramanian16}. Many of these  rely on beyond-the-standard-model physics possibly operating in the early Universe. In the post-recombination
Universe, plasma instabilities \citep[e.g.][]{Gruzinov2001, Schlickeiser2012},
the Biermann battery \citep[e.g.][]{Pudritz89, Subramanian1994, Ryu1998} and momentum transfer effects \citep[e.g.][]{Mishustin1972,Harrison1973,Saga2015} can also generate
magnetic  fields. Whether all these mechanisms are suitable for explaining the origin of the fields permeating the cosmic web is debated, and will be answered thanks to large radio telescopes (see \citealt{Beck2015} and pages 369--597 of \citealt{Bourke2015}).

We here summarise the basics of an astrophysical mechanism, based on the photoionization of the IGM, that is bound to have contributed to the magnetisation of the cosmic web during the epoch of reionisation. Its principles have been explored in \citet{Durrive2015}, and the resulting, average strength of the  field in the Universe by the end of reionisation has been estimated in \citet{Durrive2017}.

\section{Outline of the mechanism}\label{sec:proc}

The ``recipe'' for the generation of magnetic fields is, in principle, simple. First, some mechanism must spatially separate positive and negative electric charge carriers. Second, this  separation must be sustained so that  a large scale electric field is created. Third, by virtue of Faraday's law, this electric field must possess a curl.
The question of astrophysical magnetogenesis thus essentially boils down to identifying the propitious epochs and environments for such rotational electric fields to emerge.

\begin{figure}[t!]
 \centering
 \includegraphics[width=0.50\textwidth]{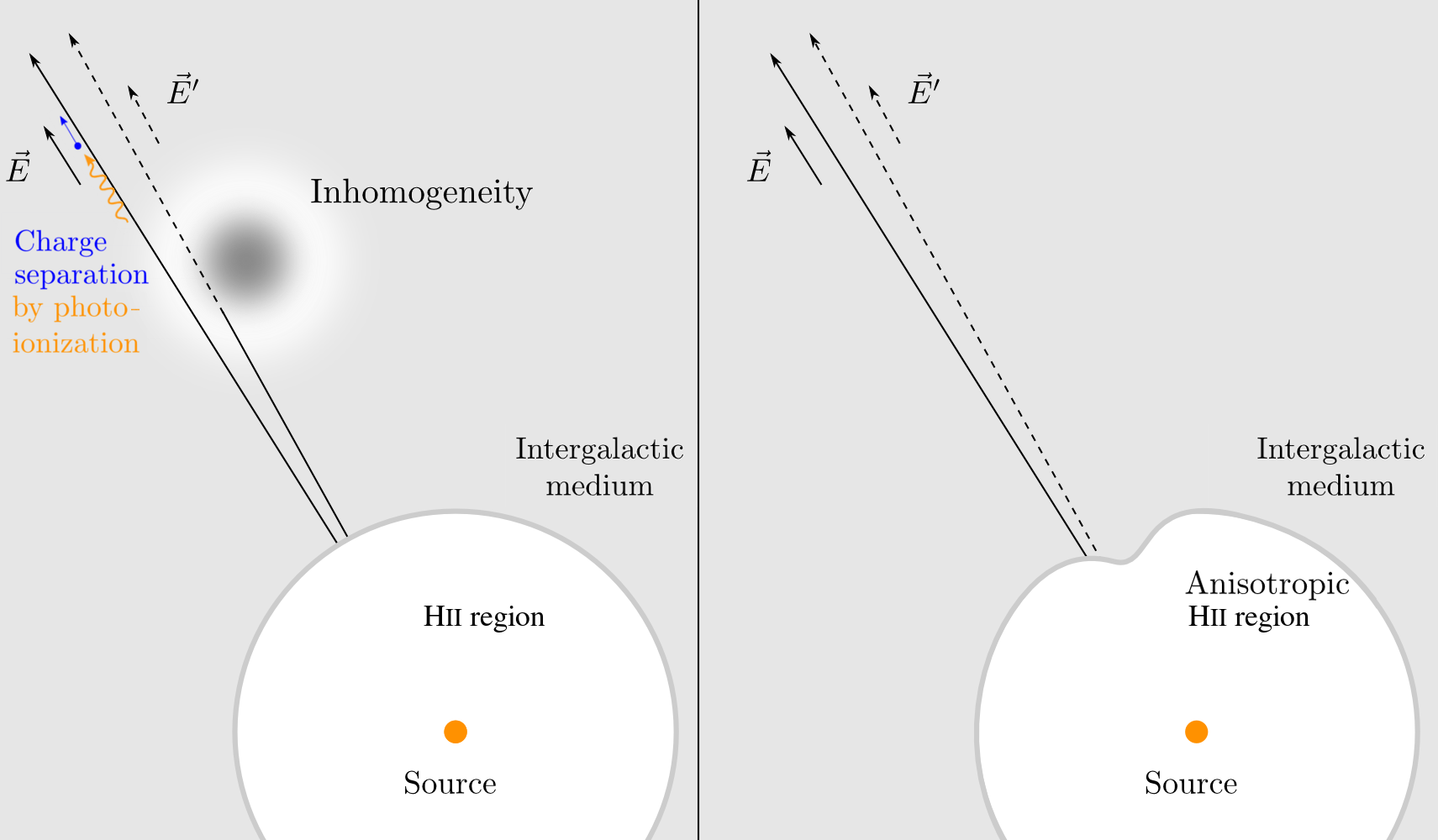}
 \caption{Inhomogeneities at the origin of rotational electric fields.
  \label{fig:aniso}}
\end{figure}

Cosmological reionisation is one such epoch. The immediate surroundings of luminous sources are ionised
forming H\textsc{ii} regions. Higher energy photons penetrate beyond the edge of such bubbles into the  neutral IGM. There, occasionally, they hit  atoms and eject new electrons. As long as the sources shine, the resulting charge separation creates an electric field. Now, in  case of perfect local isotropy, the electric field is  curl-free. However, local isotropy is broken (see fig. \ref{fig:aniso}). First, the IGM is  inhomogeneous. Any overdensity (underdensity) locally enhances (lessens) the  process: behind it, the strength of the electric field is smaller (larger) than along photon trajectories that miss density contrasts. The electric field varies \emph{across} photon trajectories, and thus possesses a curl. Second,   H\textsc{ii} bubbles are aspherical, and  the flux of hard photons that escape into the IGM is anisotropic.

In \cite{Durrive2015}, we analysed in detail this mechanism. We
obtained the  expression for the generated magnetic field, and examined  its  spatial distribution and strengths.
We considered three source types: population III stars, primeval galaxies, and quasars. We modelled a clump in the IGM by a compensated overdensity (see fig. \ref{fig:aniso} left), and assumed a source lifetime of $100$ Myr.
Population III star clusters generate relatively stronger fields, on distances ($1-2$ kpc) shorter than  half their  physical mean separation ($\sim 10$ kpc).
These sources thus leave a large fraction of the IGM unmagnetised. Rare, luminous quasars magnetise less but over much larger distances (several Mpc), comparable to half their separation.
Primeval galaxies combine modestly high amplitudes, and reasonably large scales (tens of kpc) that are similar to half their separation.

\section{Average Magnetic Energy Density seeded in the IGM}
We estimated in \cite{Durrive2017} the level of global magnetisation thus reached in the Universe by the end of reionisation. The result depends on the distribution of the ionising sources, their spectra, the epochs at which they shine, and on the density clumps in the IGM.
We used
the \citet{Press1974b} formalism  to model the statistical distribution of  sources and  overdensities.
We focused  on primeval galaxies, probably the dominant contributors to reionisation.
\begin{figure}[ht]
 \centering
 \begin{minipage}{.4\textwidth}
  \centering
  \includegraphics[width=.97\linewidth]{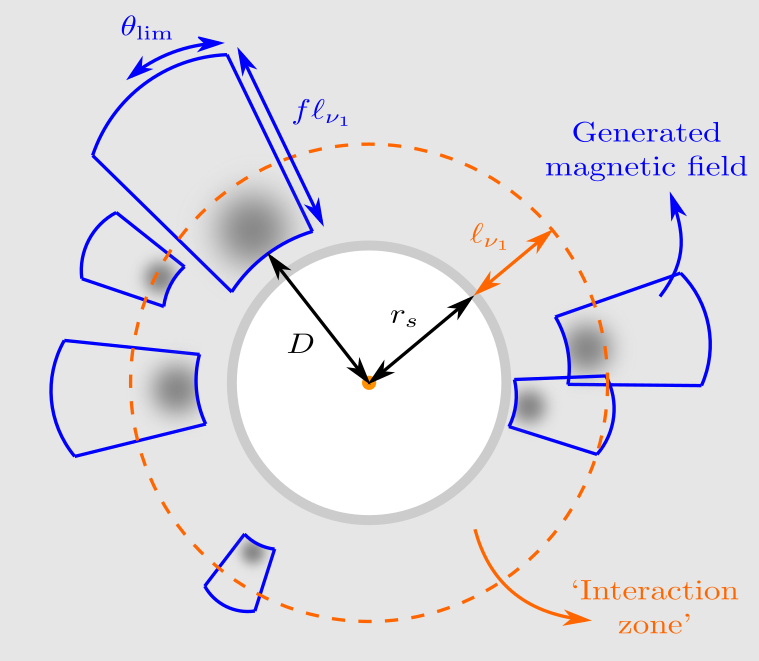}
 \end{minipage}
 \begin{minipage}{.4\textwidth}
  \centering
  \includegraphics[width=1.03\linewidth]{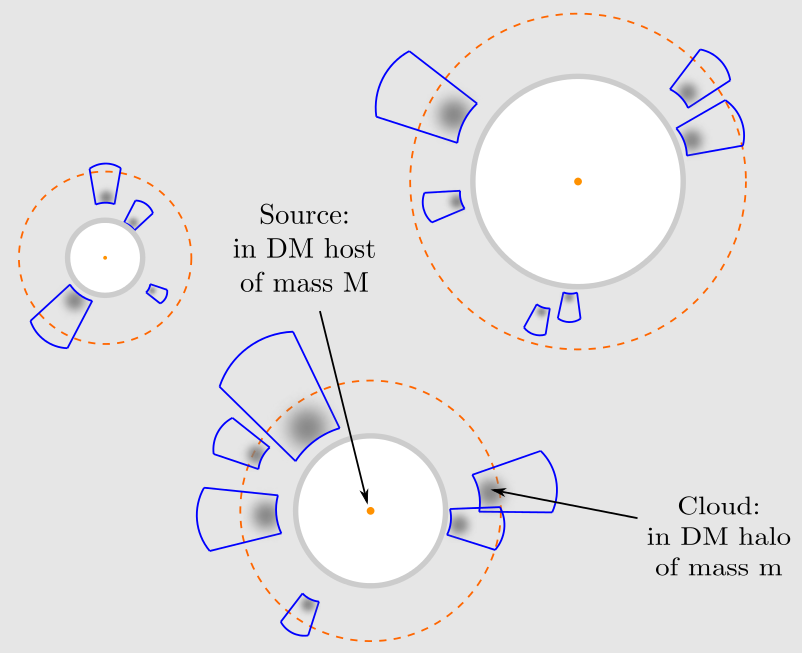}
 \end{minipage}%
 \caption{Estimating the global magnetisation level of the IGM. See text for details.}
 \label{fig:ExampleOfResultsVaryRhob}
\end{figure}

The details,  illustrated in fig. \ref{fig:ExampleOfResultsVaryRhob},  consist in the following steps:
\begin{enumerate}
 \item First, we considered an isolated source and a gas inhomogeneity in its vicinity.
       We obtained a convenient expression for the magnetic energy density $E_m(D)$ associated to any cloud of mass $m$ at a given distance $D$ of the ionising source.
 \item Second, we summed the effect of all the clouds arround the source contained in a DM halo of mass $M$. It contributes by injecting a magnetic energy
       \begin{equation}
        E_M = \int_{r_s}^{r_s + \ell_{\nu_1}} \int_{m_\mathrm{min}}^{m_\mathrm{max}}\, E_m(D)\,\dd^2 P(D, m|M)
        \label{eq:E_M}
       \end{equation}
       where $\dd^2 P(D, m|M)$ is the probability for a DM cloud of mass $m$ to be in a spherical shell of volume $4 \pi D^2 \dd D$ at  distance $D$.
       We considered only the clouds within an `interaction zone' set by the photon mean free path $\ell_{\nu_1}$ beyond which the mechanism is not  efficient.
 \item Third, we integrated the energy density $E_M$ over the DM halos containing ionising sources. As H\textsc{ii} bubbles start to overlap,
       the efficiency of magnetic field generation decreases as reionisation proceeds. Hence, we weighed the contribution of sources by a factor $1-Q_i(z)$ accounting for the ionised volume filling factor at redshift $z$. The mean comoving magnetic energy density finally reads
       \begin{equation}
        \frac{B_\mathrm{c}^2(z)}{8 \pi} = \int^{z_0}_z \dd z' \frac{1-Q_i}{H\,(1+z')^5}\int^{M_\mathrm{max}}_{M_*} \dd M\, E_M\, g_\mathrm{gl}~\frac{\dd n_M}{\dd M}
        \label{eq:b2pz}
       \end{equation}
       where $\frac{\dd n_M}{\dd M}$ is the mass function of the DM halos hosting the sources.
       The parameter $z_0$ is the redshift at which the first sources form, and $g_\mathrm{gl}$ is the rate at which sources switch on.
\end{enumerate}

\begin{figure}[ht]
 \begin{center}
  \includegraphics[width=0.50\textwidth]{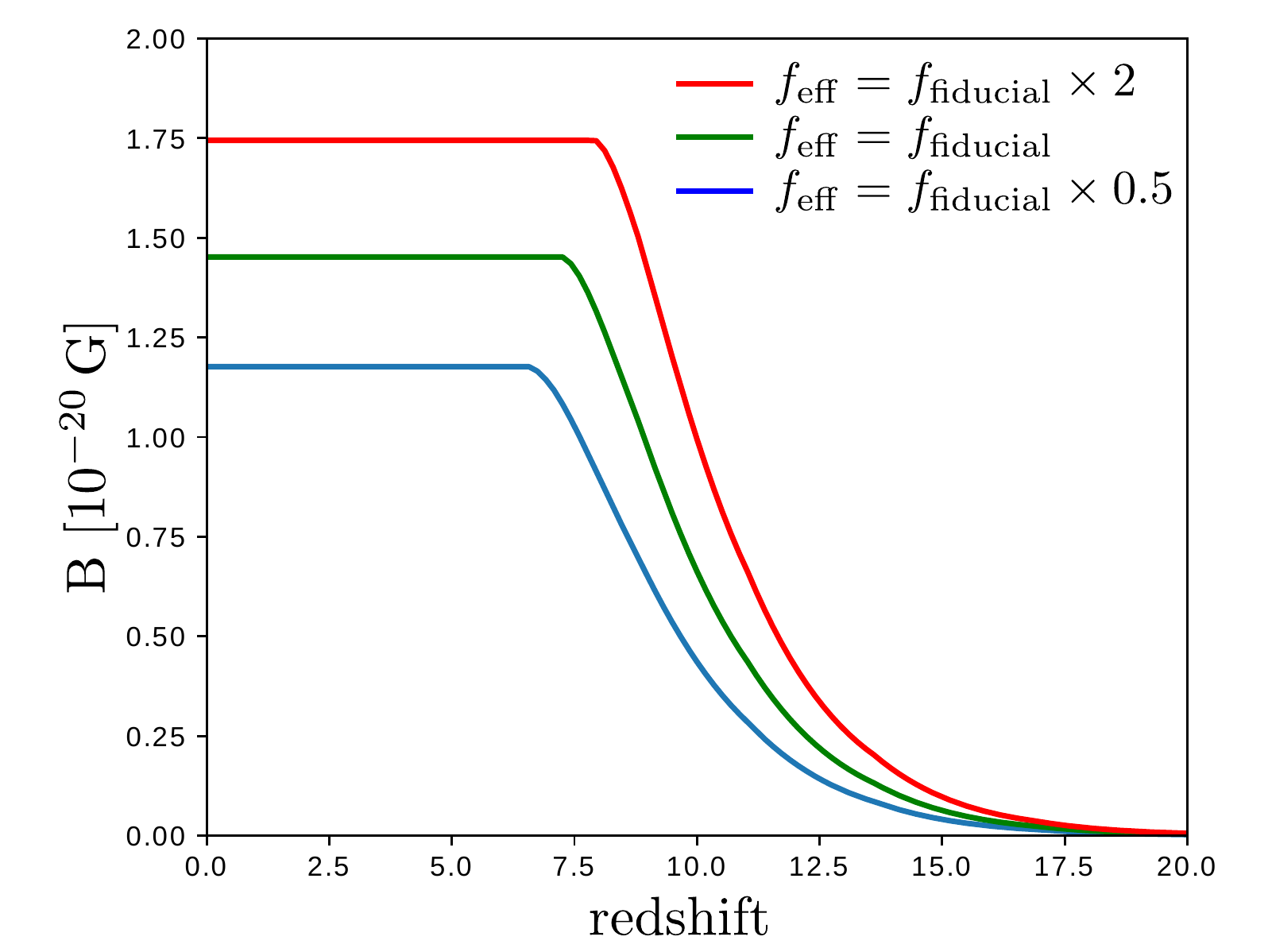}
  \caption{Evolution with redshift of the mean comoving magnetic field strength in the IGM in different reionisation histories. The green curve is the fiducial model assumed in \cite{Durrive2017} where $f_\mathrm{eff}$ is an effective reionisation efficiency.  All three considered histories  are in agreement with the \cite{PlanckCollaboration2016} constraints.}
  \label{fig:bcom}
 \end{center}
\end{figure}

Figure (\ref{fig:bcom}) shows the comoving strength of the generated magnetic field in three different reionisation histories, all consistent with results of the \cite{PlanckCollaboration2016}.
Above $z = 20$ there are no galaxies, and the magnetic field is nil. As galaxies form, their radiation induces magnetic fields that accumulate in the IGM. Once the Universe is fully ionised, the mechanism stops, and a plateau (in comoving units) is reached. Note that in physical units, the strength of the magnetic field by the end of reionisation  is a few $10^{-18}$ Gauss,  a suitable seed value for any subsequent amplification by nonlinear processes.

\section{Discussion}

The model we summarised here  can be improved in several ways. In particular,  we neglected the contribution of underdense regions, which could multiply the result obtained above by a factor of two. Similarly, we did not take into account the effect of the asphericity of the  H\textsc{ii} regions. Finally, we assumed  that the H\textsc{ii} regions have reached their steady state.
Whether taking their growing regime into account would increase or decrease the global magnetic field is not obvious. However,  nonlinearities develop in the cosmic velocity field as structure formation proceeds \citep[e.g.][]{Ryu2008,Greif2008,Sur2012}. They enter into play when the seed magnetic field has reached its final strength  \citep{Langer05}.
Magnetic field amplification  thus sets in early on, at least whithin the nodes, filaments and sheets of the cosmic web. The strength shown in fig. \ref{fig:bcom}  thus likely underestimates the actual magnetisation of those structures. In cosmic voids, plasma instabilities might have the potential to amplify rapidly the magnetic seed fields, and bring them above the lower limits suggested by the observation of blazars.

\end{document}